\documentstyle[aps,pre,multicol,epsf]{revtex}
\begin{document}
\draft
\title{ 
Persistence exponents for fluctuating interfaces}
\author{J. Krug$^1$, H. Kallabis$^2$,
S. N. Majumdar$^3$, S. J. Cornell$^4,$
A. J. Bray$^4$ and C. Sire$^5$}
\address {$^1$Fachbereich Physik, Universit\"at GH Essen, 
D-45117 Essen, Germany \\
$^2$ HLRZ, Forschungszentrum J\"ulich, D-52425 J\"ulich, Germany
\\
$^3$ Tata Institute of Fundamental Research, 
Homi Bhabha Road, Bombay 400 005, India
\\ 
$^4$ Department of Theoretical Physics, The University, 
Manchester M13 9PL, United Kingdom
\\
$^5$ Laboratoire de Physique Quantique, Universit\'e Paul Sabatier,
Toulouse, 31062 Cedex, France
}
\date{April 29, 1997}
\maketitle
\begin{abstract}
Numerical and analytic results for the exponent $\theta$ 
describing the decay of the first return probability of an interface to its 
initial height are obtained for a large class of linear Langevin equations. 
The models are parametrized  by the dynamic roughness exponent $\beta$, with 
$0 < \beta < 1$; for $\beta = 1/2$ the time evolution is Markovian. 
Using simulations of solid-on-solid models, of the discretized continuum 
equations as well as of the associated zero-dimensional
stationary Gaussian process, we  address two problems: The return of an 
initially flat interface, and the return to an initial state with
fully developed steady state roughness.
The two problems are shown to be governed by 
different exponents. For the steady state case we point out the
equivalence to fractional Brownian motion, which has a return
exponent $\theta_S = 1 - \beta$. The exponent $\theta_0$
for the flat initial condition appears to be nontrivial. We prove
that $\theta_0 \to \infty$ for $\beta \to 0$, $\theta_0 \geq \theta_S$
for $\beta < 1/2$ and $\theta_0 \leq \theta_S$ for $\beta > 1/2$,
and calculate $\theta_{0,S}$ perturbatively to first order in 
an expansion around the Markovian case $\beta = 1/2$. Using the exact
result $\theta_S = 1 - \beta$, accurate upper and lower bounds on $\theta_0$
can be derived which show, in particular, that $\theta_0 \geq 
(1 - \beta)^2/\beta$
for small $\beta$. 
\end{abstract}
\pacs{PACS numbers: 02.50.-r, 05.40.+j, 81.10.Aj}
 
\begin{multicols}{2}
\section{Introduction}

The statistics of first passage events for
non-Markovian stochastic processes has attracted
considerable recent interest in the physical 
literature. Such problems appear naturally in
spatially extended nonequilibrium systems, where the
dynamics at a given point in space becomes non-Markovian
due to the coupling to the neighbours. The asymptotic
decay of first passage probabilities turns out to 
be hard to compute even for very simple systems such as
the one-dimensional Glauber model \cite{derrida} or the
linear diffusion equation with random initial conditions
\cite{diffusion}. Indeed, determining the first passage
probability of a general Gaussian process with known
autocorrelation function is a classic unsolved problem
in probability theory \cite{slepian,reviews,majsire}.

In this paper we address the first passage statistics of 
fluctuating interfaces. The large scale behaviour of the models 
of interest is described by the linear Langevin equation
\begin{equation}
\label{Langevin}
\frac{\partial h}{\partial t} = - (- \nabla^2)^{z/2} h + \eta
\end{equation}
for the height field $h(x,t)$. Here the dynamic exponent $z$
(usually $z=2$ or 4)
characterizes the relaxation mechanism, while $\eta(x,t)$ is
a Gaussian noise term, possibly with spatial correlations.
We will generally assume a flat initial interface,
$h(x,0) = 0$. Since (\ref{Langevin}) is linear,
$h(x,t)$ is Gaussian and its
temporal statistics at an arbitrary fixed point in space is 
fully specified by the autocorrelation function 
computed from (\ref{Langevin}),
\begin{equation}
\label{auto}
A(t,t') \equiv
\langle h(x,t) h(x,t') \rangle = K [(t' + t)^{2\beta} - 
\vert t' - t \vert^{2\beta}],
\end{equation}
where $K$ is some positive constant,
and $\beta$ denotes the dynamic roughness exponent, which depends
on $z$ and on the type of noise considered. For example, for 
uncorrelated white noise $\beta = (1/2)[1 - d/z]$ for a $d$-dimensional
interface, while for volume conserving noise 
$\beta = (1/2)[1 - (d+2)/z]$ \cite{jkcam}. An interface is {\em rough}
if $\beta > 0$. In the present work we regard $\beta$ as a continuous
parameter in the interval $]0,1[$. Note that for $\beta = 1/2$ 
eq.(\ref{auto}) reduces to the autocorrelation function of a random walk,
corresponding to the limit $z \to \infty$ (no relaxation) of 
eq.(\ref{Langevin}) with uncorrelated white noise.

To define the first passage problems of interest, consider
the quantity 
\begin{equation}
\label{P(t_0,t)}
P(t_0,t) = 
{\rm Prob}[h(x,s) \neq h(x,t_0) \;\; \forall s : t_0 < s < t_0 + t] .
\end{equation}
We focus on two limiting cases. For $t_0 = 0$, $P(t_0,t)$ reduces to the
probability $p_0(t)$ that the interface has not returned to its initial
height $h=0$ at time $t$. This will be referred to as the {\em transient
persistence probability}, characterized by the exponent $\theta_0$,
\begin{equation}
\label{p0}
p_0(t) \equiv P(0,t) \sim t^{-\theta_0}, \;\;\; t \to \infty.
\end{equation}
On the other hand, for $t_0 \to \infty$ the interface develops
roughness on all scales and the memory of the flat initial 
condition is lost.
In this limit $P(t_0,t)$ describes the return to a rough
initial configuration drawn from the steady state distribution
of the process, and the corresponding {\em steady state}
persistence probability $p_S(t)$ decays with a distinct exponent $\theta_S$,
\begin{equation}
\label{ps}
p_S(t) \equiv \lim_{t_0 \to \infty} P(t_0,t) \sim t^{-\theta_S}, 
\;\;\; t \to \infty.
\end{equation} 
In general, one expects that $P(t_0,t) \sim t^{-\theta_S}$ for
$t \ll t_0$ and $P(t_0,t) \sim t^{-\theta_0}$ for $t \gg t_0$,
with a crossover function connecting the two regimes.

A particular case of the steady state persistence problem
was studied previously in the context of tracer diffusion
on surfaces \cite{jkdobbs96}. In this work it was observed
that the distribution of first return times has a natural
interpretation as a distribution of {\em trapping times} 
during which a diffusing particle is buried and cannot move;
thus the first passage exponent may translate into an anomalous
diffusion law. A simple scaling argument (to be recalled below
in Section V)  
was used to derive the relation
\begin{equation}
\label{relation}
\theta_S = 1 - \beta,
\end{equation}
which was well supported by numerical simulations for $\beta = 1/8$.
A primary motivation of the present work is therefore 
to investigate the validity
of this relation through simulations for other values of $\beta$ and
refined analytic considerations, as well as to understand why it fails
for the transient persistence exponent $\theta_0$.

The paper is organized as follows. In the next section we convert
the non-stationary stochastic process $h(x,t)$ into a stationary
Gaussian process in logarithmic time \cite{diffusion,majsire}.
This representation will provide us with a number of bounds and scaling
relations, and will be used in the simulations of Section IV.C.
A perturbative calculation of the persistence exponents in the vicinity of 
$\beta=1/2$ is presented in Section III.
The simulation results are summarized in Section IV.
Section V reviews the analytic basis of the relation 
(\ref{relation}) and makes contact to earlier work on 
the return statistics of fractional Brownian motion, while Section
VI employs the expression (\ref{relation}) for 
$\theta_S$ to numerically generate exact upper and lower bounds on 
$\theta_0$. 
Finally, some conclusions
are offered in Section VII.

\section{Mapping to a stationary process}
\label{Mapping}

Following Refs.\cite{diffusion,majsire} we introduce the normalized random 
variable $X = h/\sqrt{\langle h^2 \rangle}$ which is considered
a function of the logarithmic time $T = \ln t$. The Gaussian
process $X(T)$ is then stationary by construction,
$\langle X(T) X(T') \rangle = f_0(T - T')$, and the autocorrelation
function $f_0$ obtained from (\ref{auto}) is
\begin{equation}
\label{f0} 
f_0(T) = \cosh(T/2)^{2 \beta} - \vert \sinh(T/2) \vert^{2\beta}.
\end{equation}
In logarithmic time the power law decay (\ref{p0}) of the persistence
probability becomes exponential, $p_0(T) \sim \exp(- \theta_0 T)$,
and the task is to determine the decay rate $\theta_0$ as a functional
of the correlator $f_0$ \cite{slepian,reviews,majsire}.

Similarly a normalized stationary process can be associated 
with the steady state problem. First define the height
difference variable 
\begin{equation}
\label{H}
H(x,t;t_0) \equiv h(x,t+t_0) - h(x,t_0)
\end{equation}
and compute its autocorrelation function in the limit $t_0 \to \infty$,
\begin{eqnarray}
 A_S(t,t') &=& 
\lim_{t_0 \to \infty}
\langle H(x,t;t_0) H(x,t';t_0) \rangle\nonumber\\
&=&\lim_{t_0 \to \infty} [A(t_0+t,t_0+t') - A(t_0+t,t_0)\nonumber\\
&&\phantom{\lim_{t_0 \to \infty} [ } -A(t_0,t_0+t') + A(t_0,t_0)]\nonumber \\
\label{As}
&&=K [t^{2 \beta} + t'^{2 \beta} - \vert t' - t \vert^{2 \beta}],
\end{eqnarray}
which is precisely the correlator of fractional Brownian motion
with Hurst exponent $\beta$ \cite{brown} (see Section V).
Next $A_S(t,t')$ is normalized by $\sqrt{A_S(t,t) A_S(t',t')}$
and rewritten in terms of $T = \ln t$. This yields the 
autocorrelation function 
\begin{equation}
\label{fs}
f_S(T) = \cosh(\beta T) - 
\frac{1}{2}\vert 2 \sinh(T/2) \vert^{2\beta}.
\end{equation}

Comparison of eqs.(\ref{f0}) and (\ref{fs}) makes it plausible that the
two processes have different decay rates of their persistence
probabilities. Both functions have the same type
of short time singularity 
\begin{equation}
\label{sing}
f_{0,S}(T) = 1 - {\cal O}(\vert T \vert^{2 \beta}), \;\;\;
T \to 0,
\end{equation}
which places them in the {\em class} $\alpha = 2 \beta$ in the sense of Slepian
\cite{slepian}. However, for large $T$ they decay with different
rates, $f_{0,S}(T) \sim \exp(- \lambda_{0,S} T)$ for $T \to \infty$,
where 
\begin{equation}
\label{lambda}
\lambda_0 = 1 - \beta, \;\;\; \lambda_S = \min[\beta, 1 - \beta] 
\end{equation}
can be interpreted, in analogy with phase ordering kinetics,
as the {\em autocorrelation exponents} \cite{tjn} of the two
processes. 

For a stationary Gaussian process with a general autocorrelator $f(T)$,
the calculation of the decay exponent $\theta$ of the persistence probability
is very hard. Only in a very few cases exact results are known \cite{slepian}.
Approximate results can be derived for certain classes of
autocorrelators $f(T)$. For example, when $f(T) = 1-O(T^2)$ for small
$T$ (an example being the linear diffusion equation \cite{diffusion}),
the density of zero crossings is finite and an independent interval
approximation (IIA) \cite{diffusion} gives a very good estimate of $\theta$.
However, for any other process for which $f(T)=1- O(|T|^{\alpha})$ for
small $T$ with $\alpha < 2$, the density of zeros is infinite and the
IIA breaks down. For general proceses with $\alpha = 1$, a 
perturbative method (when the process is not far from Markovian) and
an approximate variational method was developed recently 
\cite{majsire}. This method will be applied to the present problem
in Section III.
In the remainder of this section we collect some exact bounds 
on $\theta$; further bounds will be derived in Section VI.  

Slepian \cite{slepian} has proved the following useful theorem
for stationary Gaussian processes with unit variance: 
For two processes with correlators $f_1(T)$ and $f_2(T)$ such that 
$f_1(T) \geq f_2(T) \geq 0$ for all $T$, the corresponding persistence 
probabilities satisfy $p_1(T) \geq p_2(T)$; in particular, the
inequality $\theta_1 \leq \theta_2$ holds for the asymptotic
decay rates. By applying this result to the correlators (\ref{f0})
and (\ref{fs}) we can generate a number of relations involving the
return exponents $\theta_0$ and $\theta_S$. For example, taking the
derivative of (\ref{f0}) with respect to $\beta$ one discovers
that $f_0(T)$ increases monotonically with decreasing $\beta$
for all $T$, and consequently
\begin{equation}
\label{bound0}
\theta_0(\beta) \geq \theta_0(\beta') \;\;\; 
{\rm if} \;\; \beta \leq \beta'.
\end{equation}
For $\beta \leq \beta' \leq (2 \ln 2)^{-1}$ the stronger inequality
\begin{equation}
\label{bound0b}
(1-\beta')\theta_0(\beta) \geq (1-\beta)\theta_0(\beta')
\end{equation}
is proved in the Appendix. 

Moreover, rewriting (\ref{fs}) in the form
\begin{equation}
\label{fs2}
f_S(T) = e^{- \beta \vert T \vert} + \frac{1}{2} e^{\beta \vert T \vert}
[ 1 - e^{-2 \beta \vert T \vert} - (1 - e^{- \vert T \vert})^{2 \beta}]
\end{equation}
it is evident that $f_S(T) < \exp(- \beta \vert T \vert)$ for $\beta < 1/2$
and $f_S(T) > \exp(- \beta \vert T \vert)$ for $\beta > 1/2$.
A process characterized by a purely exponential autocorrelation function
$\exp ( - \lambda \vert T \vert)$ is Markovian, and its persistence
probability can be computed explicitly \cite{slepian}; the asymptotic
decay rate $\theta$ is equal to the decay rate $\lambda$ of the correlation 
function. Thus the fact that $f_S(T)$ can be bounded by Markovian
(exponential) correlation functions supplies us with the inequalities
$$ \theta_S \geq \beta, \;\;\; \beta < 1/2 $$
\begin{equation}
\label{bounds}
\theta_S  \leq \beta, \;\;\; \beta > 1/2.
\end{equation}
The last inequality can be sharpened to 
\begin{equation}
\label{bounds2}
\theta_S  \leq 1/2, \;\;\; \beta > 1/2.
\end{equation}
This will be demonstrated in the Appendix, where we also prove
that
$$ \theta_0 \geq 1 - \beta \;\; {\rm for } \;\; \beta < 1/2 $$
\begin{equation}
\label{bound2}
\theta_0 \leq 1 - \beta \;\; {\rm for } \;\; \beta > 1/2 
\end{equation}
and
$$ \theta_0 \geq \theta_S
\;\; {\rm for } \;\; \beta < 1/2 $$
\begin{equation}
\label{bound3}
\theta_0 \leq \theta_S
\;\; {\rm for } \;\; \beta > 1/2. 
\end{equation}

Next we record some relations for special values of $\beta$.
We noted already that for $\beta = 1/2$ the interface fluctuations
reduce to a random walk, corresponding to the Markovian correlator
$f_0(T) = f_S(T) = \exp(- \vert T \vert/2)$, 
for which $\theta = \lambda = 1/2$ \cite{slepian}. Hence
\begin{equation}
\label{1/2}
\theta_0(1/2) = \theta_S(1/2) = 1/2.
\end{equation}
For $\beta = 1$ both (\ref{f0}) and (\ref{fs}) become constants,
$f_0(T) \equiv f_S(T) \equiv 1$. This implies that the corresponding
Gaussian process is time-independent, and consequently
\begin{equation}
\label{beta1}
\lim_{\beta \to 1} \theta_0(\beta) = \lim_{\beta \to 1} 
\theta_S(\beta) = 0.
\end{equation}

For $\beta \to 0$ the transient correlator (\ref{f0}) 
degenerates to the discontinuous function 
$f_0(0) = 1$, $f_0(T > 0) = 0$. 
Since this 
is bounded from above by 
the Markovian 
correlator $f(T) = e^{-\lambda \vert T \vert}$ for 
any $\lambda$, we conclude that 
\begin{equation}
\label{beta00}
\lim_{\beta \to 0} \theta_0(\beta) = \infty.
\end{equation}
In contrast, the steady state
correlator tends to a nonzero constant, $\lim_{\beta \to 0}
f_S(T) = 1/2$ for $T > 0$, with a discontinuity at $T=0$, and
therefore $\theta_S$ is expected to remain finite for 
$\beta \to 0$. 
Note that all the relations derived for $\theta_S$ -- equations
(\ref{bounds},\ref{bounds2},\ref{beta1}) -- are consistent
with $\theta_S = 1 - \beta$. 

\section{Perturbation theory near $\beta=1/2$}
\label{sec:perturb}
We have already remarked that both the steady state and the transient
processes reduce to a Markov process when $\beta=1/2$.  Two of us have
developed a perturbation theory for the persistence exponent of a
stationary Gaussian process whose correlation function is close to a
Markov process \cite{majsire}.  When the persistence probability for a
Markov process is written in the form of a path integral, it is found
to be related to the partition function of a quantum harmonic oscillator 
with a hard wall at the origin.  The persistence probability for a process 
whose correlation function differs perturbatively from the Markov process, 
i.e. whose autocorrelation function is
\begin{equation}
\label{firstf}
f(T) = \exp\left(-\lambda |T|\right) + \epsilon \phi(T), 
\end{equation}
may then be calculated from a knowledge of the eigenstates of the
quantum harmonic oscillator.  In (\ref{firstf}) we have used the 
same normalization, $f(0)=1$, as elsewhere in this paper. With this 
normalization,  $\phi(0)=0$. (Note that a different normalization was 
employed in ref.\ \cite{majsire}.)  

If $\beta=1/2 + \epsilon$, equations (\ref{f0}) and (\ref{fs}) may be
written in the form
\begin{equation}
f_{0,S}=\exp\left( -{|T|\over 2}\right) + 
\epsilon\phi_{0,S}(|T|) + {\cal O}(\epsilon^2),
\end{equation}
where
\begin{eqnarray}
\phi_0&=&2\cosh {T\over 2}\ln\left(\cosh {T\over 2}\right)-2\sinh
{T\over 2}\ln\left(\sinh {T\over 2}\right),\label{phi0}\\
\phi_S&=&\sinh {T\over 2}\left(T-2\ln\left(2\sinh {T\over 2}\right)\right).
\label{phis}
\end{eqnarray}

The result for the persistence exponent (equivalent to equation (7) of 
reference \cite{majsire}) may most conveniently be written in the 
form \cite{Klaus}
\begin{equation}
\theta=\lambda\left\{1-\epsilon
{2\lambda\over\pi}\int_0^\infty\phi(T)\left[1- \exp (-2\lambda
T)\right]^{-3/2} dT\right\}.
\label{pertform}
\end{equation}
Substituting (\ref{phi0}) and (\ref{phis}) into (\ref{pertform}), one
finds (after some algebra)
\begin{eqnarray}
\theta_0&=&{1\over 2}-\epsilon(2\sqrt{2}-1)+
{\cal O}(\epsilon^2),\label{theta0}\\
\theta_S&=&{1\over 2}-\epsilon +{\cal O}(\epsilon^2).\label{thetas}
\end{eqnarray}
Eqn.\ (\ref{thetas}) agrees with the relation $\theta_S=1-\beta$,
while (\ref{theta0}) compares favourably with the
stationary Gaussian process simulations for $\beta=0.45$ and $\beta=0.55$ 
to be presented in Section IV.C.

\section{Simulation results}
\label{Simulations}

\subsection{Solid-on-solid models}

Simulations of one-dimensional, 
discrete solid-on-solid models were carried out for
$\beta = 1/8$, $\beta = 1/4$ and $\beta = 3/8$. The case $\beta = 1/8$
describes an equilibrium surface which relaxes through surface diffusion,
corresponding to $z=4$ in (\ref{Langevin}) and volume conserving noise with
correlator
\begin{equation}
\label{noisec}
\langle \eta(x,t) \eta(x',t') \rangle = - \nabla^2 \delta(x - x') 
\delta(t - t').
\end{equation}
The cases $\beta = 1/4$ and $\beta = 3/8$ are realized for nonconserved
white noise in (\ref{Langevin}) and dynamic exponents $z=2$ and $z=4$,
respectively \cite{jkcam}.

In all models the interface configuration is described by a set of
integer height variables $h_i$ defined on a one-dimensional lattice
$i = 1,...,L$ with periodic boundary conditions. For simulations of 
the transient return problem, large lattices ($L = 2 \times 10^5$ --
$2 \times 10^6$) were used, while for the steady state problem
we chose small sizes, $L = 100-200$ for dynamic exponent
$z=4$ and $L = 1000$ for $z=2$, in order
to be able to reach the steady state within the simulation time. 

The precise simulation procedure is somewhat dependent on whether
the volume enclosed by the interface is conserved 
(as for $\beta = 1/8$) or not. To simulate the transient return
problem with conserved dynamics, the interface was prepared in
the initial state $h_i = 0$, and each site was equipped with a counter
that recorded whether the height $h_i$ had returned to $h_i = 0$.
The fraction of counters still in their initial state then gives the
persistence probability $p_0(t)$. For the steady state problem the interface
was first equilibrated for a time $t_{\rm eq}$ large compared to the 
relaxation time $\sim L^z$ \cite{jkcam,relax}. Then the configuration
$h_i(t_{\rm eq})$ was saved, and the fraction of sites which had
not yet returned to $h_i(t_{\rm eq})$ was recorded over a prescribed
time interval $t_{\rm eq} < t < t_{\rm eq} + t_1$. At the end of that
interval the current configuration $h_i(t_{\rm eq}+t_1)$ was chosen 
as the new initial condition, and the procedure was repeated.
After a suitable number of repetitions (typically $10^4$ for $z=4$,
and 2000 for $z=2$), the surviving fraction gives
an estimate of $p_S$.

The models used in the cases $\beta = 1/4$ and $\beta = 3/8$ are growth
models, in which an elementary step consists in chosing at random a site
$i$ and then placing a new particle,
$h_j \to h_j +1$, either at $j = i$ or at one of the
two nearest neighbour sites $j = i \pm 1$, depending on the local environment.
For these nonconserved models the procedures described above have to be
modified such that the calculation of the surviving fractions
$p_0$ and $p_S$ is performed only when a whole monolayer -- that is,
one particle per site -- has been deposited. At these instances the
average height is an integer which can be subtracted from the whole
configuration in order to decide whether a given height variable
has returned to its initial state when viewed in a frame moving with
the average growth rate.
 
We now briefly describe the results obtained in the conserved case.
In \cite{jkdobbs96} the steady state return problem 
for $\beta = 1/8$ was investigated
in the framework of the standard one-dimensional solid-on-solid model
with Hamiltonian 
\begin{equation}
\label{SOS}
{\cal H} = J \sum_i \vert h_{i+1} - h_i \vert 
\end{equation}
and Arrhenius-type surface diffusion dynamics \cite{jkwagner}. We have
extended these simulations to longer times and to different values of
the coupling constant $J$. Figure 1 shows that the exponent $\theta_S$
is independent of $J$, and that its value is numerically indistinguishable
from $\theta_S = 7/8$ predicted by (\ref{relation}).

Since the transient persistence probability $p_0$ decays very rapidly for
$\beta = 1/8$, a more efficient model was needed in order to obtain 
reasonable statistics. We therefore used a
restricted solid-on-solid (RSOS) model introduced by R\'acz et al.
\cite{racz}. In this model the nearest neighbour height differences
are restricted to 
\begin{equation}
\label{rsos}
\vert h_{i+1} - h_i \vert \leq 2.
\end{equation}
In one simulation step a site $i$ is chosen at random, and a diffusion
move to a randomly chosen neighbour is attempted. If the attempt fails
due to the condition (\ref{rsos}), a new random site is picked. Figure
2 shows the transient persistence probability $p_0(t)$ obtained from a 
large scale simulation of this model. The curve still shows considerable
curvature, and we are only able to conclude that probably $\theta_0 > 3.3$
for this process.

\begin{figure}[b]
\narrowtext
\epsfxsize=\hsize
\epsfbox{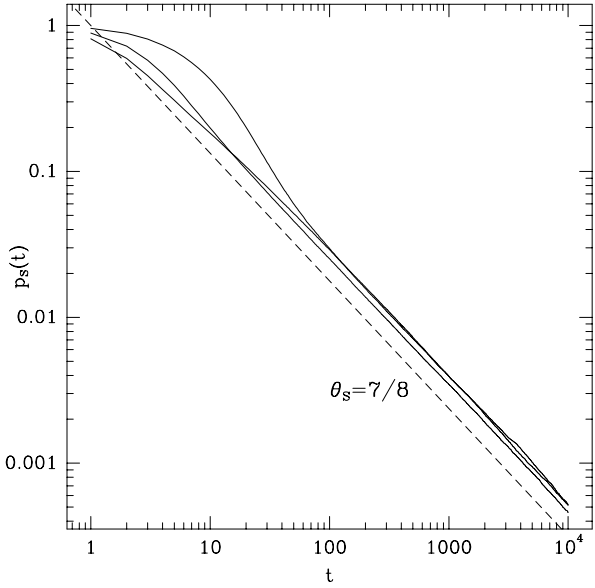}
\caption{ Steady state persistence probability $p_S(t)$ for the Arrhenius
surface diffusion model with coupling constant $J = 0.25$, 0.5 and 1. 
Systems of size
$L = 100$ were equilibrated for $t_{\rm eq} = 5 \times 10^7$ attempted
 moves per site.
Then data were collected over $10^4$ time intervals of length $t_1 = 10^4$. 
The dashed
line has slope $\theta_S = 1 - \beta = 7/8$.
}
\end{figure}
\begin{figure}[b]
\narrowtext
\epsfxsize=\hsize
\epsfbox{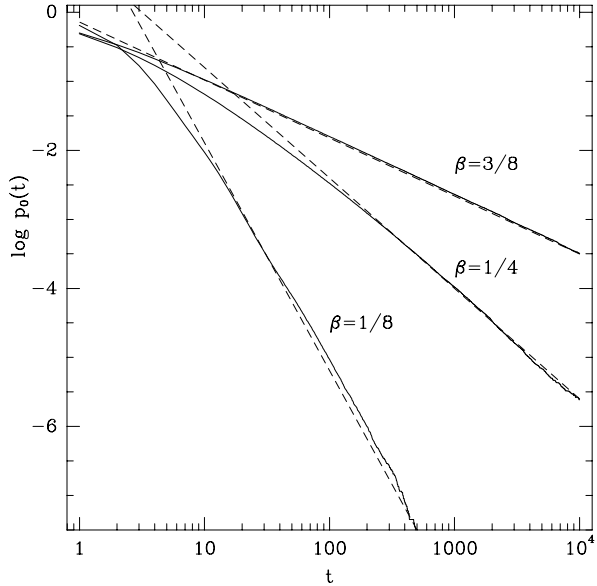}
\caption{Transient persistence probability $p_0(t)$ for three 
different solid-on-solid
models described in the text. The system sizes used were $L = 2 \times 10^5$ 
for $\beta = 1/8$,
and $L = 2 \times 10^6$ for $\beta = 1/4$ and $\beta = 3/8$. For $\beta = 1/8$
 ($\beta = 1/4$)
an average over 1000 (10) runs was taken, while the data for $\beta = 3/8$ 
constitute a single
run. The slopes of the dashed lines correspond to the exponent estimates 
in the fourth column
of Table I. 
}
\end{figure}

Figure 2 also shows transient results for $\beta = 1/4$ and 
$\beta = 3/8$. In the former case we used a growth model introduced
by Family \cite{fam86}, in which the deposited particle is always
placed at the lowest among the chosen site $i$ and its neighbours,
whereas for $\beta = 3/8$ we used the curvature model introduced 
in \cite{jk94}. Our best estimates of $\theta_0$ for these models
are collected in Table I, along with the values for $\theta_S$
which agree, within numerical uncertainties, with the relation 
(\ref{relation}) in all cases.

\subsection{Discretized Langevin equations}

We solved equation (\ref{Langevin}) in discretized time and space
for the real valued function $h(x_i,t_n)$, where
$t_n \equiv n\Delta t$ and $x_i\equiv i\Delta x$ with $n = 0, 1, 2, \dots$ and
$i=0, \dots, L-1$ in a system with periodic boundary conditions.

For the time discretization we used a simple forward Euler
differencing scheme \cite{NumRec}:
\begin{equation}
\frac{\partial h(x_i, t_n)}{\partial t} \equiv \frac{h(x_i,t_{n+1}) -
  h(x_i,t_n)}{\Delta t}.
\end{equation}
The spatial derivatives for the cases $z=2$ and $z=4$ considered
in the simulations were discretized as
\begin{eqnarray}
\nabla^2 h(x_i) & \equiv & h(x_{i-1}) - 2 h(x_i) + h(x_{i+1}),\\
(\nabla^2)^2 h(x_i) & \equiv & h(x_{i-2}) -4 h(x_{i-1}) + 6 h(x_{i})\nonumber\\
&& - 4 h(x_{i+1}) + h(x_{i+2})
\end{eqnarray}
for the function $h(x_i)\equiv h(x_i,t_n)$ at any given time $t_n$.
Here and in the simulations, 
the spatial lattice constant $\Delta x$ was set to unity.

With these definitions, we iterated the equation
\begin{eqnarray}
h(x_i,t_{n+1}) &=& h(x_i,t_n) - \Delta t(-\nabla^2)^{z/2} h(x_i, t_n) 
\nonumber\\
&& \phantom{ h(x_i,t_n) } + \sigma \sqrt{\Delta t} \eta(x_i, t_n)
\label{DiscLangevin}
\end{eqnarray}
where $\eta(x_i,t_n)$ is a Gaussian distributed random number with
zero mean and unit variance whose correlations will be specified below.

Von Neumann stability analysis \cite{NumRec} shows that values
$\Delta t \leq 1/2$ and 1/8 for $z=$ 2 and 4, respectively, have to be
used to keep the noise--free iteration stable.
The simulations showed that the
scheme remained stable with $\Delta t=0.4$ and 0.1 even in the
noisy case.

\begin{figure}
\narrowtext
\epsfxsize=\hsize
\epsfbox{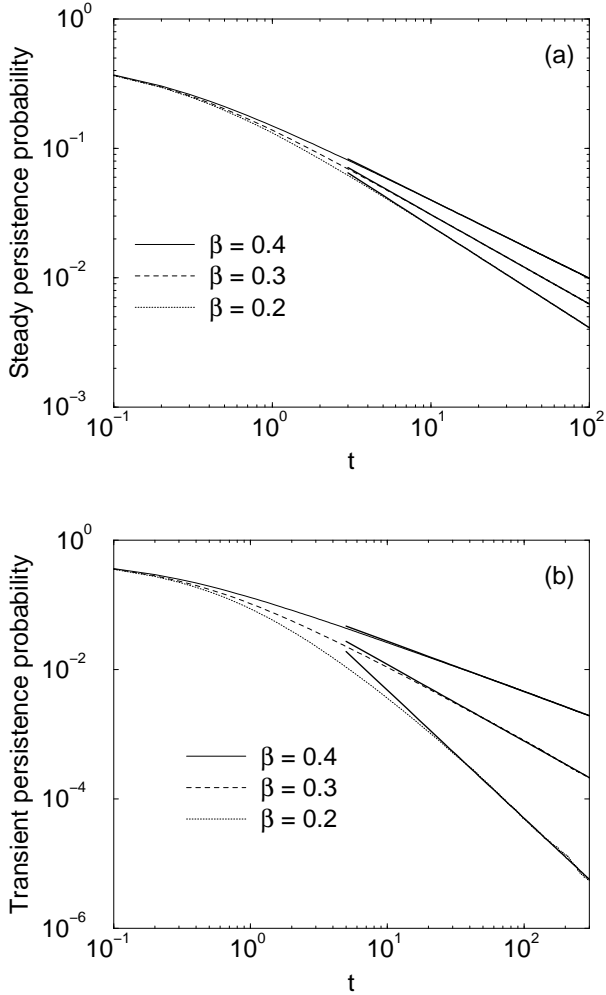}
\caption{(a) Steady state persistence probability $p_S(t)$ from the 
numerical solution of 
the discretized Langevin equation with $z=2$ and correlated noise 
($\rho = -0.1, 0.1, 0.3$). The
thick lines represent fits to the last decade of $p_S(t)$. The slopes 
are given in the
second column of Table I. (b) Same as (a) for the transient persistence 
probability $p_0(t)$. 
}
\end{figure}

We used white noise $\eta_w(x_i,t_n)$ with a correlator
\begin{equation}
\langle \eta_w(x_i,t_n) \eta_w(x_j,t_m)\rangle 
        = \delta_{i,j} \delta_{n,m}
\end{equation}
in the simulations,
as well as spatially correlated noise $\eta_\rho(x_i,t_n)$ with 
\begin{equation}
\langle \eta_\rho(x_i,t_n)\eta_\rho(x_j,t_m)\rangle 
  = \sigma^2 g_\rho(x_i-x_j)\delta_{n,m}
\end{equation}
where 
\begin{equation}
g_\rho(x_i-x_j)\equiv 
\left\{\begin{array}{rl}
|x_i-x_j|^{2\rho-1}&, i \ne j\\
1 & , i=j
\end{array} \right. 
\end{equation}
and $\rho<1/2$ a real number.
A different choice of regularizing $g(0)$ \cite{HH} 
did not change the results. 
If $S_\rho(k)$ denotes the discrete Fourier transform of
$g_\rho$, we defined
\begin{equation}
\label{corrnoise}
\eta_\rho(k,t)\equiv \sqrt{|S_\rho(k)|}r_k\exp(2\pi i\varphi_k)
\end{equation}
with $r_k$ being a Gaussian distributed amplitude with zero mean and
$\varphi_k\in[0, 1[$ a uniformly distributed random phase \cite{HH}.
Due to the regularization $g(0)=1$, which fixes the average value of
$S_\rho$, one has to use the modulus in (\ref{corrnoise}) as 
$S_\rho$ can be negative for some $k$.
Iterating eqn.\ (\ref{DiscLangevin}) with $z=2$, the correlated noise
(\ref{corrnoise}) with $\rho < 1/2$ leads to surface
roughness with a measured roughness exponent $\beta_m$ 
that agrees with the prediction $\beta=(1+2\rho)/4$
of the continuum equation (\ref{Langevin}) within error bars.
The case $\beta > 1/2 $, i.e.\ $\rho>1/2$ is not accessible by this method.

The simulated systems had a size of $L=4096$, noise strength
$\sigma=1$ and averages were typically taken over 3000 independent runs.

In all cases, the simulation was started with a flat inital condition
$h(x,0)=0$.
To measure the persistence probabilities, the configuration $h(x,t_0)$ 
and the consecutive one $h(x,t_0+\Delta t)$ were kept in memory during the
simulation. In each following iteration $t_n=(t_0+\Delta t) + n\Delta t$,
$n=1,2,3,\dots$, an initially zeroed counter at each site $x$ was 
increased as long 
as $\mbox{sgn}(h(x,t_n)-h(x,t_0))=\mbox{sgn}(h(x,t_0+\Delta t)-h(x,t_0))$. 
The fraction of counters with a value larger than $t$ then gave the
persistence probability $p(t)$. For measurement of $\theta_0$, $t_0$ was chosen
to be zero, for $\theta_S$, $\Delta t\ll t_0 \ll L^z$, and the power
law behaviour in the regime $\Delta t \ll t \ll t_0$ was used.

For comparison, $\theta_S$ was also measured in small systems $L=128$
in the steady state $t\gg L^z$ for $z=2$ with uncorrelated noise.
The results agreed with the measurements for $\Delta t\ll t_0 \ll L^z$.
However, in the steady state, one has to take care of measuring
the power law
decay of $p_S(t)$ only up to the correlation time $\sim L^z$, so that
we preferred measurement in the regime $\Delta t\ll t_0 \ll L^z$ here.

Figure 3 shows typical curves $p_S(t)$ and
$p_0(t)$ obtained from
the numerical solution of the discretized Langevin equation
(\ref{DiscLangevin}) with correlated noise for the values $\rho=-0.1$,
0.1 and 0.3.
A summary of all measured persistence exponents $\theta_0$ and $\theta_S$ as 
a function of the roughness exponent $\beta$ can be found in Table I.

\subsection{Simulation of the stationary Gaussian process}
\label{Sim:process}

Since a Gaussian process is completely specified by its correlation
function, it is possible to simulate it by
constructing a time series that possesses the same correlation
function.  This is most easily performed in the frequency domain.

Suppose $\tilde \eta(\omega)$ is (the Fourier transform of) a Gaussian
white noise, with 
$\langle\tilde\eta(\omega)\tilde\eta(\omega')\rangle=2\pi
\delta(\omega+\omega')$. 
Let
\begin{equation}
\tilde X(\omega)=\tilde\eta(\omega)\sqrt{\tilde f(\omega)}, 
\end{equation}
where $\tilde f(\omega)$ is the Fourier transform of the desired
correlation function (notice that $\tilde f(\omega)$ is the power
spectrum of the process, so it must be positive for all
$\omega$).  Then the correlation function of $\tilde X$ is
\begin{equation}
\langle\tilde X(\omega)\tilde X(\omega')\rangle = 2\pi\tilde f(\omega)
\delta(\omega+\omega').
\end{equation}
The inverse Fourier transform $X(T)$ is therefore a stationary
Gaussian process with correlation function $\langle
X(T)X(T')\rangle=f(T-T')$.  

The simulations were performed by constructing Gaussian (pseudo-)white noise
directly in the frequency domain, normalizing by the appropriate
$\sqrt{\tilde f(\omega)}$, then fast-Fourier-transforming back to the
time domain.  The distribution of intervals between zeros, and hence
the persistence probability, may then be measured directly. 
The resulting process $X(T)$ is periodic, 
but this is not expected to affect the results provided
the period $N\delta T$ is sufficiently long, where $N$ is the number
of lattice sites used and $\delta T$ is the time increment between the
lattice sites.  It is desirable for $N$ to be as large as possible,
consistent with computer memory limitations---typically $N=2^{19}$ or $2^{20}$.
The timestep $\delta T$ must be sufficiently small for the short-time
behaviour of the correlation function to be correctly simulated, but
also sufficiently large that the period of the process is not too
small.  Typical values for $\delta T$ were in the range
$10^{-4}$--$10^{-2}$.  Several different values of $\delta T$ were
used for each $\beta$, to check for consistency, and the results were
in each case averaged over several thousand independent samples.

This method works best for processes that are ``smooth''.  The density of
zeros for a stationary Gaussian process is $\sqrt{-f''(0)}/\pi$
\cite{rice}, which is only finite  if
$f(T)=1-{\cal O}(T^2)$.   However, the correlation functions
for the processes under
consideration behave like (\ref{sing}) at short time, so they have an
infinite density of zeros.  Any finite discretization scheme will
therefore necessarily miss out on a large (!) number of zeros, and
will presumably overestimate the persistence probability and hence
underestimate
$\theta$ (this drawback is not present if Eqn. (\ref{Langevin}) is simulated
directly, since $\delta T=\delta t/t$ effectively becomes zero as
$t\to\infty$).   Nevertheless, the simulations were found to be
well-behaved when $\beta$ was greater than 0.5,
with consistent values of
$\theta$ for different values of $\delta T$.  However, when  $\beta$
was less than 0.5 it
was found to be increasingly difficult to observe such convergence
before finite size effects became apparent.  Convergence was not
achieved for 
$\beta<0.45$.
Figure 4 shows the persistence probability for both the steady state 
and the transient processes with $\beta=0.45$ and $\beta=0.75$, 
for two values of $\delta
T$ in each case. The agreement of the data for the different values of
$\delta T$ is better for the larger value of $\beta$.  
\begin{figure}
\narrowtext
\epsfxsize=\hsize
\epsfbox{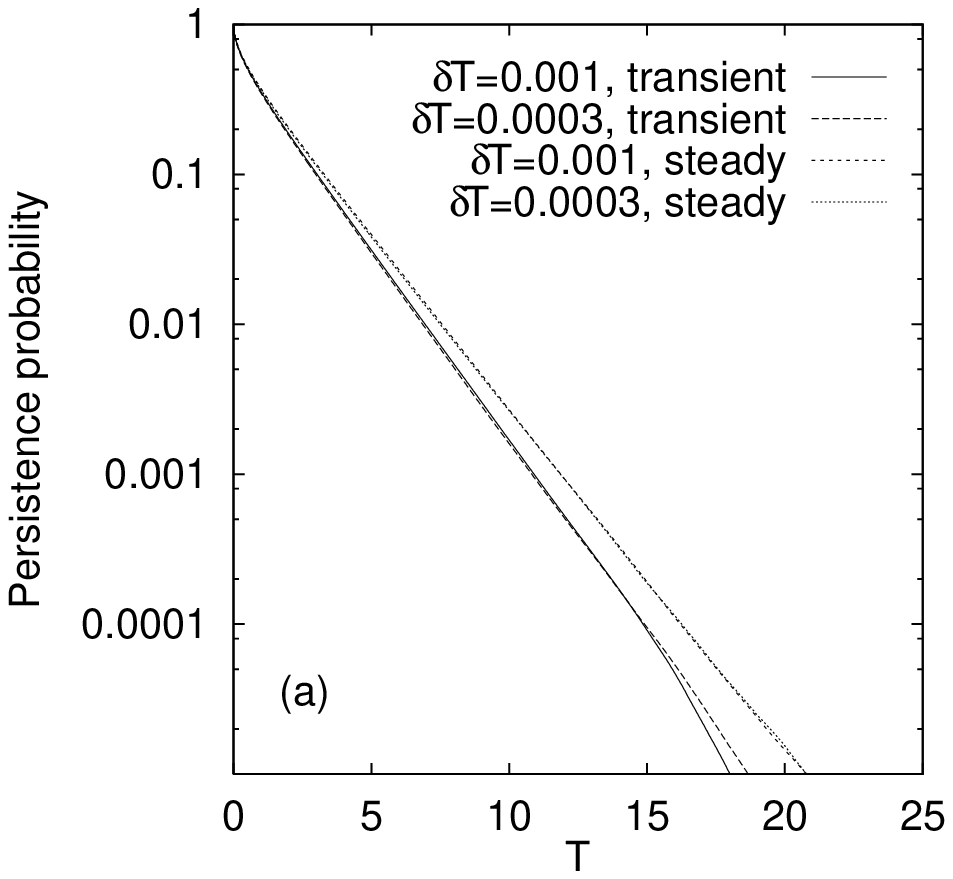}
\epsfxsize=\hsize
\epsfbox{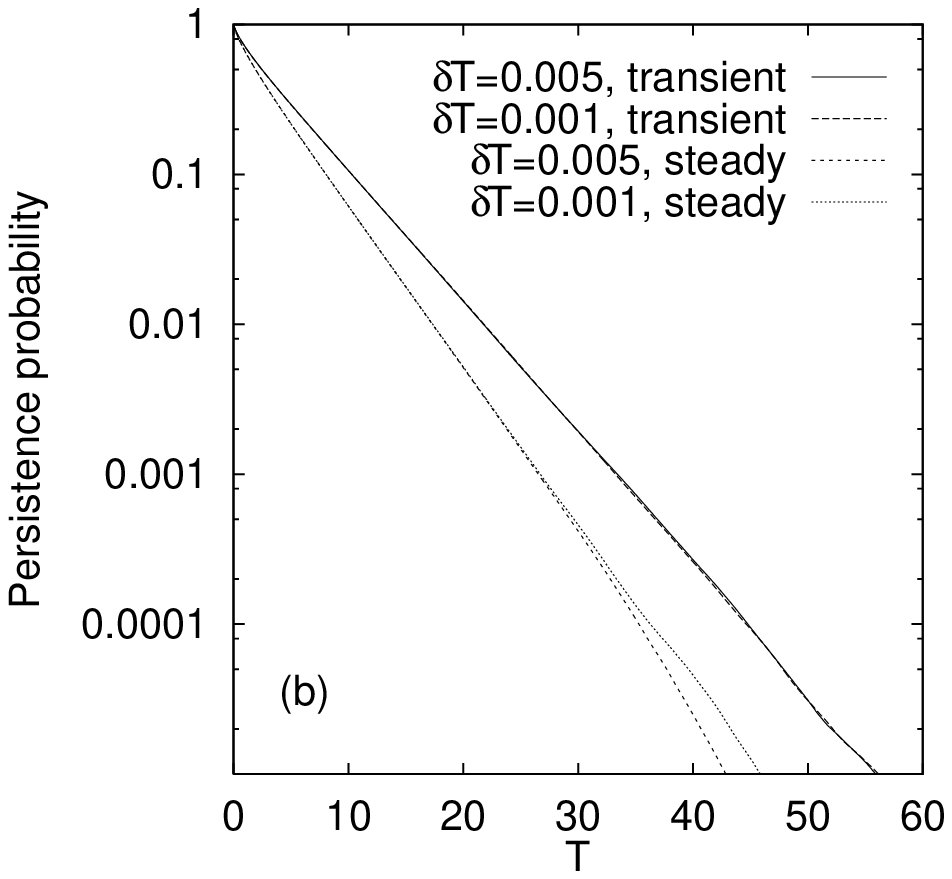}
\caption{(a) Persistence probability obtained from simulations of
 the equivalent stationary
Gaussian process for both the transient and the steady-state case, with 
$\beta = 0.45$. Two
values of the time increment $\delta T$ are shown. (b) Same as (a) for 
$\beta = 0.75$.
}
\end{figure}

A summary of the measured values of $\theta$ for different values of
$\beta$ is to be found in Table I.   The quoted errors
are subjective estimates based on the consistency of the results
for different values of $\delta T$, and are smaller for the larger
values of $\beta$ due to the process being ``smoother''. For
$\beta>0.5$, the estimated 
values of $\theta_S$ agree well with the prediction
$1-\beta$, while the result for $\theta_0$ when $\beta=0.55$ agrees
well with the perturbative prediction $\approx 0.409$ from Eqn.\
(\ref{theta0}). For
$\beta=0.45$, the prediction $\theta_S=1-\beta$ lies outside the
quoted error bars, reflecting the difficulty of assessing the
systematic errors in the simulations, while the result for $\theta_0$
is consistent with the prediction $\approx 0.591$ of the perturbation theory.

\section{The persistence exponent of fractional Brownian motion}
\label{Scaling}

The numerical results presented in the preceding section,
as well as the pertubative calculation of Section III, 
clearly
demonstrate the validity of the identity (\ref{relation}),
$\theta_S = 1 - \beta$,
over the whole range $0 < \beta < 1$. Since the special property
of the steady state process which is responsible
for this simple result is obscured by the mapping 
performed in Section II, we now return to the
original, unscaled process in time $t$. The crucial observation
is that the limiting process $\lim_{t_0 \to \infty} H(x,t;t_0)$ 
defined in (\ref{H},\ref{As}) has,
by construction, {\em stationary increments}, in the sense that
\begin{eqnarray}
\sigma^2(t,t') &\equiv &\lim_{t_0 \to \infty} \langle (H(x,t; t_0) - 
H(x,t'; t_0))^2 \rangle\nonumber\\
& =& 2 K \vert t - t' \vert^{2 \beta}
\label{incr}
\end{eqnarray}
depends only on $t - t'$. The power law behaviour of the variance
(\ref{incr})  
is the defining property of the {\em fractional Brownian motion}
(fBm)
introduced by Mandelbrot and van Ness \cite{brown}, and identifies
$\beta$ as the `Hurst exponent' of the process. 

The first return statistics of fBm has been
addressed previously in the literature 
\cite{berman,hansen,ding}, and analytic
arguments as well as numerical simulations supporting the relation 
(\ref{relation}) have been presented.
It seems that the relation in fact applies more
broadly, to general self-affine processes which need
not be Gaussian \cite{hansen,maslov}. For completeness 
we provide in the following a simple derivation along the lines
of \cite{hansen,ding,maslov}. 

We use $H(t)$ as a short hand for the fBm limit of $H(x,t; t_0)$ for
$t_0 \to \infty$. Let $H_1 \equiv H(t_1)$, and define $\rho(\tau)$ as
the probability that $H$ has returned to the level $H_1$ (not necessarily
for the first time) at time $t_1 + \tau$. Obviously, using (\ref{incr}) we
have
\begin{equation}
\label{rho}
\rho (\tau) = \frac{1}{\sqrt{2 \pi} \sigma(\tau)} \sim 
\tau^{-\beta}, \;\;\; \tau \to \infty.
\end{equation}
The set of level crossings becomes sparser with increasing distance
from any given crossing. It is `fractal' in the sense that the density,
viewed from a point on the set, decays as $\tau^{-(1-D)}$ with 
$D = 1 - \beta$. 

We now relate the decay (\ref{rho}) to that
of the persistence probability $p_S(t)$. Consider a time interval
of length $L \gg 1$. According to (\ref{rho}), 
the total number $N(L)$ of returns to the level
$H_1 = H(t_1)$ in the interval $t_1 < t < t_1 + L$ is of the order
\begin{equation}
\label{N(L)}
N(L) \sim L^{1 - \beta}.
\end{equation}
Now let 
\begin{equation}
\label{q}
q(\tau) \sim - dp_S/d\tau \sim \tau^{-(1+\theta_S)}
\end{equation}
denote the probability distribution of time intervals between level crossings.
The number $n(\tau)$ of intervals of length $\tau$ within the interval
$[ t_1, t_1 + L]$
is proportional to $q(\tau)$, and can be written as
\begin{equation}
\label{n}
n(\tau) = n_0(L) \tau^{-(1+\theta_S)},
\end{equation}
where the prefactor $n_0(L)$ is fixed by the requirement that the total
length of all intervals should equal $L$, i.e. 
\begin{equation}
\label{totalL}
\int_0^{L} d\tau \; \tau n(\tau) \sim L.
\end{equation}
This gives $n_0(L) \sim L^{\theta_S}$, and since $N(L) \sim n_0(L)$
comparison with (\ref{N(L)}) yields the desired relation (\ref{relation}).

In \cite{jkdobbs96} an essentially equivalent argument was given, however
it was also assumed that the intervals between crossings are independent,
which is clearly not true due to the strongly non-Markovian character
of the fBm \cite{brown}. Here we see that what is
required is not independence, but only stationarity of interval
spacings (that is, of increments of $H$). The latter property does not 
hold for the transient problem, where the probability of an
interval between subsequent crossings depends not only on its length,
but also on its position on the time axis. 
For the transient process the variance of temporal increments
can be written in a form analogous to (\ref{incr}),
\begin{equation}
\label{transvar}
\sigma^2(t,t') = \langle (h(x,t) - 
h(x,t'))^2 \rangle = 2 K a_\beta(t'/t) \vert t - t' \vert^{2 \beta},
\end{equation}
where the positive function $a_\beta$ interpolates monotonically 
between the limiting values 
$a_\beta(0) = 2^{2 \beta - 1}$ and $a_\beta(1) = 1$. This would appear 
to be a rather
benign (scale-invariant) `deformation' of the fBm, however, as we have seen,
the effects on the persistence probability are rather dramatic
for small $\beta$. 
Similar considerations may apply to the Riemann-Liouville fractional Brownian
motion \cite{brown}, which shares the same kind of 
temporal inhomogeneity \cite{lim}.

It is worth pointing out that {\em stationary} Gaussian 
processes with a short time singularity
$f(T) \sim 1 - {\cal O}(\vert T \vert^{\alpha})$, $T \to 0$
(compare to (\ref{sing})) have level crossing sets which
are fractal, in the mathematical sense \cite{falconer}, 
with Hausdorff dimension
$D = 1 - \alpha/2$ \cite{orey,marcus}.
However, as
was emphasized by Barb\'e \cite{barbe}, this result only describes
the short time structure of the process; a suitably defined scale-dependent
dimension of the coarse-grained set
always tends to unity for large scales, since the coarse-grained
density of crossings is finite. In other words, the crucial relation
(\ref{rho}) holds for $\tau \to 0$ but {\em not} 
for $\tau \to \infty$. In the present context this implies that 
although the stationary
correlators $f_0$ and $f_S$ share the same short-time singularity
(eq.(\ref{sing})), for $f_0$ this does not provide us with any
information about the persistence exponent $\theta_0$.

\begin{figure}
\narrowtext
\epsfxsize=\hsize
\epsfbox{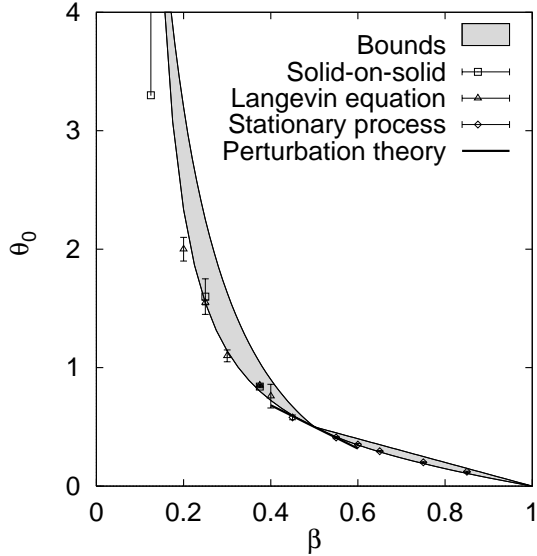}
\caption{Summary of data for the transient exponent $\theta_0(\beta)$, 
obtained from
simulations of the solid-on-solid models (squares), discretized Langevin
 equations (triangles)
and the equivalent stationary process (diamonds). The bold line is the 
perturbation result
(\protect{\ref{theta0}}), while the shaded area 
shows the range enclosed by the exact
 upper and lower
bounds derived in Section VI.
}
\end{figure}

\section{Exact numerical bounds for $\theta_0$}
\label{Newbounds}
In Section \ref{Mapping} several rigorous analytic bounds for $\theta_0$
were obtained by comparing the correlator $f_0(T)$ to a function with
known persistence exponent and using Slepian's theorem \cite{slepian}.
Having convinced ourselves, in the preceding section, that the expression
(\ref{relation}) for $\theta_S$ is in fact exact in the whole interval
$0 < \beta < 1$, we can now employ the same strategy to obtain further
bounds by comparing $f_0(T,\beta)$ 
to $f_S(T,\beta)$. These bounds turn out to be very
powerful because $f_0(T,\beta)$ and $f_S(T,\beta)$ 
share the same type of singularity
at $T=0$ (see eqn.(\ref{sing})). 

Let $f_A(T)$ and $f_B(T)$ be monotonically decreasing functions 
with the same class of analytical behaviour near $T=0$.  There will 
always exist numbers $b_{\rm min}$ and $b_{\rm max}$ such that 
$f_A(b_{\rm max} T)\le f_B(T)\le f_A(b_{\rm min} T)$ is satisfied 
for all $T$.  Since the persistence exponent for a process with correlation 
function $f_A(b T)$ is just $b \theta_A$, where $\theta_A$ is the 
persistence exponent for the process with correlation function $f_A(T)$, we can
deduce from Slepian's theorem that $b_{\rm\min}
\theta_A\le\theta_B\le b_{\rm\max}\theta_A$.  We may therefore find 
upper and lower bounds for the persistence exponent of a given process if we 
know the persistence exponent for another process in the same class.  In 
particular, we may use the result (\ref{relation})
for $\theta_S$ to obtain bounds for
$\theta_0$.

Finding by analytic means the most restrictive values of $b_{\rm min}$ 
and $b_{\rm max}$ such that $f_S(b_{\rm max} T,\beta)\le f_0
(T,\beta)\le f_S (b_{\rm min} T, \beta)$ is a formidable task.  
Our approach is to 
study the analytical behaviour near $T=0$ and $T=\infty$ to find values 
where the inequalities are satisfied in the vicinity of both limits, and 
then investigate numerically whether the inequalities are satisfied away 
from these asymptotic regimes.  The same leading small-$T$ behaviour
is obtained for $f_S(b T, \beta)$ and $f_0(T,\beta)$ when
$b=2^{{1\over 2\beta}-1}$, and the same large-$T$ behaviour is
found when $b=(1-\beta)/\beta$ for $\beta<1/2$ and $b=1$
for $\beta>1/2$.  We conclude that
\begin{eqnarray}
b_{\rm min}&\le& \left\{ 
\begin{array}{l c}
\hbox{min}\left( 2^{{1\over 2\beta}-1}, {1\over\beta} -1\right) & 
\hbox{ for $\beta<1/2$}\\
2^{{1\over 2\beta}-1} & \hbox{ for $\beta>1/2$}
\end{array}
\right.
\label{amin}
\\
b_{\rm max}&\ge& \left\{ 
\begin{array}{l c}
\hbox{max}\left( 2^{{1\over 2\beta}-1}, {1\over\beta} -1\right) & 
\hbox{ for $\beta<1/2$}\\
1 & \hbox{ for $\beta>1/2$}
\end{array}
\right.
\label{amax}
\end{eqnarray}
Surprisingly, we find in the majority of cases that the inequalities
(\ref{amin}) and 
(\ref{amax}) are satisfied as equalities.  Specifically, in the
cases where the corrections to the leading analytic behaviour has the 
appropriate sign to ensure that $f_S(b T, \beta)$ is a bound for
$f_0(T, \beta)$ in both limits $T\to0$ and $T\to\infty$, then
numerical investigation revealed it to be a bound for all $T$.  We are
therefore able in the following cases 
to quote the most restrictive bounds and their region
of applicability in analytical form, although the validity of the
bounds has only been established numerically.  
\begin{eqnarray}
\theta_0 & \ge & {(1-\beta)^2\over\beta}  \qquad\hbox{ for 
$0<\beta<\beta_1(=0.1366\dots)$}\label{tmin1}\\
\theta_0 &\ge& (1-\beta)2^{{1\over 2\beta}-1} \qquad\hbox{ for
$\beta_1<\beta<{1\over 2}$}\label{tmin2}\\
\theta_0 & \le &{(1-\beta)^2\over\beta}  \qquad\hbox{ for 
$\beta_2(=0.1549\dots)<\beta<{1\over 2}$}\label{tmax1}\\
\theta_0 & \le & 1-\beta \qquad\hbox{ for 
$\beta>{1\over 2}$}\label{tmax2}
\end{eqnarray}
The two critical values $\beta_1=0.1366\dots$ and
$\beta_2=0.1549\dots$
 correspond to 
the solutions in $]0, 0.5[$ of ${1\over\beta_1}-1=2^{{1\over 2\beta_1}-1}$ 
and $\beta_2=2^{2\beta_2-3}$ respectively.  Note that (\ref{tmax2})
coincides with the rigorous result in Eq. (\ref{bound2})

For other values of $\beta$, the
sign of the leading corrections implies that (\ref{amin}) and 
(\ref{amax}) is only
satisfied as an inequality, and the best
value of the bound has to be obtained numerically by finding the
value of $b$ where  $f_S(b T, \beta)$ touches $f_0(T,
\beta)$ at a point.

It is also possible to obtain bounds on $\theta_0(\beta)$ by
comparing $f_0(T, \beta)$ with $f_S(b T, \beta')$, where
$\beta'\ne\beta$.  Consideration of the behaviour at small-$T$ shows
that a lower bound on $\theta_0$ may be
obtained when $\beta'>\beta$, 
whereas an upper bound may be obtained for $\beta'<\beta$.
In the majority of cases, it can be shown that the most
restrictive bounds are in fact obtained by $\beta'=\beta$.  For
instance, consideration of the large-$T$ behaviour for $0<\beta<\beta_1$
shows that $b_{\rm min}\le{1\over\beta'}-1$, so $\theta_{\rm
  min}\le\left( {1\over\beta'}-1\right)(1-\beta).$  Therefore, since
this inequality is satisfied as an equality for $\beta'=\beta$ (see 
(\ref{tmin1})),  and
$\beta'\ge\beta$, the best bound is obtained when
$\beta'=\beta$.  Similarly, the inequalities (\ref{tmax1}) and (\ref{tmax2})
can be shown to be the best obtainable by this method.
However, for $\beta_1<\beta<1/2$, a perturbative consideration with
$\beta'=\beta+\epsilon$ shows that a larger value of $\theta_{\rm
min}$ may be obtained for $\epsilon$ small and positive, so
(\ref{tmin2}) is not the best bound that may be obtained.  A
numerical investigation shows that the optimal value of $\beta'$ is
nevertheless very close to $\beta$, and an improvement in $\theta_{\rm min}$
of no more than
$\approx 3\%$ was so obtained.  For $\beta$ outside the ranges quoted in 
(\ref{tmin1}--\ref{tmax2}),  it was found by numerical investigation
that $\beta=\beta'$ gave the best bound.

Numerical values of these bounds are listed in Table~I, for
comparison with the simulation data.  The contents of the last three columns of
this table are
also plotted in Figure 5.  The upper and lower bounds are 
(perhaps) surprisingly close together.  Notice that the lower bound for 
$\theta_0$ when $\beta=0.125$ is $6.125$, whereas the discrete solid-on-solid
simulations yielded
the inconclusive value $>3.3$.  All the other data are 
consistent with the bounds, within numerical error.  
It is interesting to note that the data, as well
as the exact perturbation theory result, tend to lie
much closer to the lower than the upper bound.

\section{Conclusions}
\label{Conclusions}

In this paper we have investigated 
the first passage statistics for a one-parameter family of 
non-Markovian, Gaussian stochastic processes which arise in the
context of interface motion. We have identified two persistence
exponents describing the short time (transient)
and long time (steady state) regimes, respectively. 

For the steady
state exponent the previously conjectured relation $\theta_S = 
1 - \beta$ \cite{jkdobbs96} was confirmed. While this relation
follows from simple scaling arguments applied to the original
process, it is rather surprising when viewed from the perspective
of the equivalent stationary process with autocorrelation function
(\ref{fs}): It provides the {\em exact} decay exponent for a family
of correlators whose class $\alpha = 2 \beta$ 
(in the sense of Slepian \cite{slepian})
covers the whole interval $\alpha \in ]0,2[$; previous
exact results were restricted to $\alpha = 1$ and
$\alpha = 2$ \cite{slepian}. 
We have demonstrated in Section VI how
this can be exploited to obtain accurate
upper and lower bounds for other processes within the same class.   

Estimates for the nontrivial transient exponent $\theta_0$
were obtained using a variety of analytic, exact and perturbative
approaches, as well as from simulations. The numerical techniques --
direct simulation of interface models and construction of realizations
of the equivalent stochastic process, respectively -- are complementary,
in the sense that the former is restricted to the regime $\beta < 1/2$,
while the latter gives the most accurate results for $\beta > 1/2$.
The results summarized in Figure 5 provide a rather complete picture
of the function $\theta_0(\beta)$.

Finally, we briefly comment on a possible experimental realization of our work.
Langevin equations of the type (\ref{Langevin}) are now widely used
to describe time-dependent step fluctuations on crystal surfaces
observed with the scanning tunneling microscope
\cite{bartelt}. From such measurements
the autocorrelation function of the
step position can be extracted, 
and different values of $\beta$ have been observed,
reflecting different dominant mass transport mechanisms
\cite{steps}. Thus it seems that,
perhaps with a slight refinement of the observation techniques, the first
passage statistics of a fluctuating step may also be accessible to 
experiments. 

\subsection*{Acknowledgements}
The authors gratefully acknowledge the hospitality of ICTP,
Trieste, during a workshop at which this research was initiated.
J.K. wishes to thank Alex Hansen for comments, and
the DFG for support within SFB 237 {\em Unordnung und
grosse Fluktuationen}. 
Laboratoire de Physique Quantique is Unit\'e Mixte de Recherche C5626 of
Centre National de la Recherche Scientifique (CNRS).

\section*{Appendix: Derivation of some exponent inequalities}

\def\theequation {A\arabic{equation}}
\setcounter{equation}{0}

To establish (\ref{bounds2}) we need to show that $f_S(T) \geq 
\exp(- \vert T \vert/2)$
for all $T$, provided that $\beta > 1/2$. To this end we rewrite 
(\ref{fs}) in the form
\begin{equation}
f_S(T) = \frac{1}{2} e^{\beta \vert T \vert} 
[1 + e^{-2 \beta \vert T \vert} - (1 - e^{- \vert T \vert})^{2 \beta}]
\end{equation}
and notice that for $\beta > 1/2$, $[1 - \exp(- \vert T \vert)]^{2 \beta} \leq
1 - \exp(- \vert T \vert)$. Thus
\begin{equation}
f_S(T) \geq \frac{1}{2} [ e^{-\beta \vert T \vert} + 
e^{-(1 - \beta) \vert T \vert}] \geq 
e^{- \vert T \vert/2},
\end{equation}
where the last inequality follows from the fact that the expression
in the square brackets is an increasing function of $\beta$ for 
$\beta > 1/2$.

Next we consider the relations (\ref{bound2}). We express (\ref{f0})
in the form 
\begin{equation}
\label{f02}
f_0(T) = 2^{-2 \beta} e^{\beta \vert T \vert} g(e^{- \vert T \vert})
\end{equation}
where the function $g$ is given by 
\begin{equation}
g(y) = (1 + y)^{2 \beta} - (1 - y)^{2 \beta}.
\end{equation}
Taking two derivatives with respect to $y$ it is seen that
$g'' \geq 0$ for $\beta < 1/2$ and $g'' \leq 0$ for $\beta > 1/2$.
Since $g(0) = 0$ and $g(1) = 2^{2 \beta}$ always, it follows that
$g(y)$ is bounded by the linear function $2^{2 \beta} y$, from above
for $\beta < 1/2$ and from below for $\beta > 1/2$. Inserted back
into (\ref{f02}) this implies 
$$
f_0(T) \leq e^{-(1 - \beta) \vert T \vert} \;\; {\rm for} \;\;
\beta < 1/2 $$
\begin{equation}
f_0(T) \geq e^{-(1 - \beta) \vert T \vert} \;\; {\rm for} \;\;
\beta > 1/2,
\end{equation}
and (\ref{bound2}) follows by applying Slepian's theorem 
\cite{slepian} in conjunction with the fact that $\theta = \lambda$
for purely exponential (Markovian) correlators.  

The inequalities (\ref{bound3}) are a little more
subtle to prove. Let us first consider the case ${1/2}< \beta < 1 $.
We need to prove that $f_0(T) \geq f_S(T)$ for all $T$. Then
the relation ${\theta}_0 \leq {\theta}_S$ will follow from Slepian's
theorem. Denoting $y=e^{-|T|}$ and using the expressions of $f_0(T)$ and
$f_S(T)$, we then need to prove that the function $F(y)=(1+y)^{2\beta}
+(a-1)(1-y)^{2\beta}-a(y^{2\beta}+1)$ (where $a=2^{2\beta-1}$) is
positive for all $0\leq y \leq 1$. 

First we note, by simple Taylor
expansion around $y=0$ and $y=1$, that $F(y) > 0$ for $y$ close to
$0$ and $1$. The first derivative, $F'(y)$ starts at the positive
value $2\beta(2-a)$ at $y=0$ and approaches $0$ from the negative side as 
$y\to 1^{-}$. The second derivative $F''(y)$ starts at $-\infty$ as $y \to
0^{+}$ and approaches $+\infty$ as $y \to 1^{-}$. We first show that
$F''(y)$ is a monotonically increasing function of $y$ in $0\leq y \leq 1$.

To establish this, we consider the third derivative, $F'''(y)=
2\beta (2\beta-1)(2\beta -2)G(y)$, where
\begin{equation}
G(y)= (1+y)^{2\beta-3}-(a-1)(1-y)^{2\beta-3}-ay^{2\beta-3}.
\end{equation}
Now, since $(1+y)^{(2\beta-3)} < y^{(2\beta-3)}$ for all $0\leq y \leq 1$,
we have $G(y)< -(a-1)[y^{2\beta-3}+ (1-y)^{2\beta -3}]$ implying
$G(y) < 0$ for $0\leq y \leq 1$. Since $1/2 \leq \beta \leq 1$, it follows
that $F'''(y) > 0$ for all $0\leq y \leq 1$. Therefore, $F''(y)$ is a
monotonically increasing function of $y$ for $0\leq y \leq 1$ and hence
crosses zero only once in the interval $[0,1]$. This implies that the
first derivative $F'(y)$ has one single extremum in $[0,1]$. However,
since $F'(y)$ starts from a positive value at $y=0$ and approaches $0$
from the negative side as $y\to 1^{-}$, this single extremum must be a
minimum. Therefore, $F'(y)$ crosses zero only once in $[0,1]$ implying
that the function $F(y)$ has only a single extremum in $[0,1]$. Since,
$F(y)$ for $y\to 0^{+}$ and $y\to 1^{-}$, this must be a maximum.
Furthermore, $F(y)$ can not cross zero in $[0,1]$ because that would
imply more than one extremum which is ruled out. This therefore
proves that $f_0(T) \geq f_S(T)$ for all $T$ for $\beta \geq 1/2$
and hence, using Slepian's theorem, $\theta_0 \leq \theta_S$.
Using similar arguments, it is easy to see that the reverse, $\theta_0\geq
\theta_S$ is true for $\beta \leq 1/2$. 

Finally we prove the inequality (\ref{bound0b}) which
relates the values of $\theta_0$ for two different exponents
$\beta$ and $\beta' > \beta$, subject to an additional constraint
to be specified below. In
the same spirit as above, one can show that after defining
$\gamma =(1-\beta)/(1-\beta')>1$, one obtains
\begin{eqnarray}
\label{ineqgen}
f_0(T,\beta) &&\leq f_0(\gamma T,\beta')\\ 
&&\hbox{ for }
\beta < \beta'\hbox{ and } 
\beta 2^{-2\beta} \leq \beta' 2^{-2\beta'}\nonumber.
\end{eqnarray}
Both functions in (\ref{ineqgen}) decay exponentially at large time with the
same decay rate $\lambda_0=1-\beta=\gamma(1-\beta')$, such that their ratio
approaches a constant when $T\to \infty$. The last condition  $\beta
2^{-2\beta}\leq\beta' 2^{-2\beta'}$ expresses that the limit of this ratio
must be less than unity. 
As $f_0(T,\beta) \leq f_0(\gamma T,\beta')$ in the
vicinity of $T=0$ (this is just a consequence of $\beta<\beta'$), 
and as a careful
study shows that $\frac{d}{dT}[f_0(T,\beta)/f_0(\gamma T,\beta')]$ has at
most one zero in the range $]0,\infty[$, we conclude that
$f_0(T,\beta)/f_0(\gamma T,\beta')<1$ if and only if the limit of this ratio
for $T\to\infty$ is less than unity. In practice, the last condition in
(\ref{ineqgen}) expressing this constraint can be violated only if
$\beta > 1/2$ and $\beta' > (2 \ln 2)^{-1} \approx 0.7213475...$.

Using Slepian's theorem, and the fact that 
the persistence exponent associated with
$f_0(\gamma T,\beta)$ is exactly $\gamma$ times the persistence 
exponent associated
with $f_0(T,\beta)$ \cite{slepian}, we arrive at the inequality
(\ref{bound0b}) which holds under the conditions stated in (\ref{ineqgen}).
Setting $\beta$ or $\beta'$ equal to $1/2$ (keeping $\beta <
\beta'$), eq.(\ref{bound0b}) reduces to the bounds (\ref{bound2}). 

For $\beta' > \beta > 1/2$ the inequality (\ref{bound0b}) comes rather
close to being satisfied as an equality. For example, setting
$\beta = 0.55$, $\beta' = 0.85$ eq.(\ref{bound0b}) requires that
$\theta_0(\beta) / \theta_0(\beta') \geq 3$, while the numerical
data in Table I yield $\theta_0(\beta) / \theta_0(\beta') = 3.39$. 
The only pair of values in Table I which violates the inequality
(\ref{bound0b}) is $(\beta, \beta') = (0.75,0.85)$. Since
for these values the condition $\beta
2^{-2\beta}\leq\beta' 2^{-2\beta'}$ is also violated, this may
be taken as an indication that the numerical estimates for $\beta > 1/2$
are rather accurate. 

\end{multicols}
\widetext
\vspace{2 cm}
\begin{center}
\begin{tabular}{|l||ll|l||ll|l|l|} \hline
\hfil$\beta$ &&$\theta_S$ & $1-\beta$&&$\theta_0$&$\theta_{\rm min}$ &
$\theta_{\rm max}$ \\ \hline
0.125$^{(*)}$ & $0.86$&$\pm 0.02$ & 0.875 &   $ > 3.3 $&& 6.125 & 7.359\\\hline
0.2  & $0.788$&$ \pm 0.01$ & 0.8 & $2.0$&$ \pm 0.1$ & 2.333 & 3.200 \\ \hline
0.25$^{(*)}$ & $0.754$&$\pm 0.01$& 0.75 & $ 1.6$&$\pm 0.15$ & 1.547 & 2.250 \\
0.25  & $0.740$&$\pm 0.01$& 0.75 & $ 1.55$&$\pm0.02$ & 1.547 & 2.250 \\ \hline
0.3   & $0.69$&$\pm 0.01$ & 0.7 &$1.10$&$\pm 0.05$ & 1.141 & 1.633\\ \hline
0.375$^{(*)}$ & $0.635$&$\pm 0.01$ & 0.625 & $0.84$&$\pm 0.01$& 0.801& 1.042 \\
0.375 & $0.625$&$\pm 0.01$ & 0.625 & $ 0.85$&$\pm 0.01$ &0.801&1.042\\ \hline
0.4 & $0.60$&$\pm 0.01$ & 0.6 & $0.76$&$\pm 0.1$ & 0.723 & 0.900  \\ \hline
0.45 & $0.53$&$\pm 0.01$ & 0.55 & $0.58$&$\pm 0.02$ & 0.598 & 0.672 \\ \hline
0.55 & $0.44$&$\pm 0.01$ & 0.45 & $0.41$&$\pm 0.01$ & 0.415 & 0.450 \\ \hline
0.6 & $0.39$&$\pm 0.01$ & 0.4 & $0.35$&$\pm 0.01$ &0.348 & 0.400 \\ \hline
0.65 & $0.346$&$\pm 0.005$ & 0.35 & $0.295$&$\pm 0.005$ & 0.289&0.350\\ \hline
0.75 & $0.247$&$\pm 0.005$ & 0.25&$0.201$&$\pm 0.005$ & 0.191& 0.250 \\ \hline
0.85 & $0.150$&$\pm 0.005$ & 0.15&$0.121$&$\pm 0.005$ &0.107 & 0.150 \\ \hline
\end{tabular}
\end{center}
\vspace{0.5cm}
\noindent
{\bf Table I.} Numerical estimates of the
persistence exponents $\theta_S$ and $\theta_0$, compared to the
fractional Brownian motion result $\theta_S = 1 - \beta$ (third column) and
to the optimal bounds
$\theta_{\rm min}$ and $\theta_{\rm max}$ derived in Section VI (last two
columns).
With the exception of the
values marked with an asterisk$^{(*)}$, which were obtained using
discrete solid-on-solid models (Section IV.A),
the data for $\beta \leq 0.40$
are taken from simulations of discretized Langevin equations
(Section IV.B) while those for $\beta \geq 0.45$ were generated using
the equivalent stationary Gaussian process (Section IV.C). In all
cases the error bars reflect a subjective estimate of systematic
errors.


\begin{thebibliography}{99}
\bibitem{relax} For a realistic estimate of the required relaxation
time it is useful to know also the prefactor of the relation
$t_{\rm eq} \sim L^z$. This is available analytically for the Arrhenius
model ($\beta = 1/8$) 
\cite{jkwagner} and numerically for the curvature model 
($\beta = 3/8$) \cite{jkcam}. In the steady state simulations reported
here we chose $t_{\rm eq} = 5 \times 10^7$ for the $z=4$--models 
(system size $L=100$), and $t_{\rm eq} = 2 \times 10^7$ for the 
$z=2$--model (system size $L=1000$).
\end{thebibliography}
\end{document}